\documentclass[preprint,12pt]{article}

\usepackage{color}
\usepackage{mathtools}
\usepackage{multicol}
\usepackage{graphicx}
\usepackage{amssymb}
\usepackage{amsmath}
\usepackage{cases}
\usepackage{booktabs,multirow}
\usepackage{soul}
\usepackage{color}
\usepackage[linesnumbered,ruled]{algorithm2e}
\usepackage{url}
\usepackage{scrextend}

\newcommand{\tL}{\mathcal{L}}

\newtheorem{rem}{Remark}

\begin{document}


\begin{center}
\textbf{\large Electric Vehicle -- Wireless Charging-Discharging Lane Decentralized Peer-to-Peer Energy Trading} \\[10pt]
   {\large Dinh Hoa Nguyen} \\[6pt]
  International Institute for Carbon-Neutral Energy Research (WPI-I$^2$CNER), \\ [6pt]
	Institute of Mathematics for Industry (IMI), \\[6pt]
  Kyushu University, Fukuoka, Japan  \\[6pt]
  Email: hoa.nd@i2cner.kyushu-u.ac.jp
\end{center}

\begin{abstract}
This paper investigates the problem of bidirectional energy exchange between electric vehicles (EVs) and road lanes embedded with wireless power transfer technologies called wireless charging-discharging lanes (WCDLs). As such, EVs could provide better services to the grid, especially for balancing the supply-demand, while bringing convenience for EV users, because no cables and EV stops are needed. To enable this EV--WCDL energy exchange, a novel decentralized peer-to-peer (P2P) trading mechanism is proposed, in which EVs directly negotiate with a WCDL to reach consensus on the energy price and amounts to be traded. Those energy price and amounts are solutions of an optimization problem aiming at optimizing private cost functions of EVs and WCDL. The negotiation process between EVs and WCDL is secured by a privacy-preserving consensus mechanism. Further, to assure successful trading with desired energy price and amounts, an analytical and systematic method is proposed to select cost function parameters by EVs and WCDL in a fully decentralized manner. Simulations are then carried out to validate developed theoretical results, which confirm the effectiveness and scalability of the proposed algorithm.
\end{abstract}


\newpage

\section*{Nomenclature}
\begin{labeling}{Lowerupper bounds}
\item[P2P] Peer-to-peer.
\item[EV] Electric vehicle.
\item[WPT] Wireless power transfer.
\item[IWPT] Inductive wireless power transfer.
\item[WCDL] Wireless charging-discharging lane.
\item[DR] Demand response.
\item[MAS] Multi-agent system.
\item [$E_{{\rm V},i}$, $E_{{\rm L},i}$, $E_{{\rm L}}$] Trading energy of EV $i$, of WCDL with EV $i$, and total trading energy of WCDL [kWh].
\item[$f_{{\rm V},i}$, $f_{{\rm L}}$] Private cost functions of EV $i$ and WCDL.
\item[$a_{{\rm V},i}$, $b_{{\rm V},i}$] Parameters in the cost function of EV $i$. 
\item[$a_{{\rm L}}$, $b_{{\rm L}}$] Parameters in the cost function of WCDL. 
\item[$\underline{E}_{{\rm V},i}$, $\overline{E}_{{\rm V},i}$] Lower and upper bounds for trading energy of EV $i$. 
\item[$\underline{E}_{{\rm L}}$, $\overline{E}_{{\rm L}}$] Lower and upper bounds for trading energy of WCDL. 
\end{labeling}

\section{Introduction}

Recent disasters worldwide as aftermaths of global climate changes, which greatly affect to human living and global economics, urge serious actions worldwide, in which reducing greenhouse gas (GHG) emissions is a must. To achieve that, transportation and energy systems should be on priority since their portions in the total GHG emissions are highest among industrial sectors. 

Currently, transportation systems are increasingly being electrified, while energy systems including electric power grids are witnessing a rapid transformation from fossil fuel based and centralized generation to renewable based and increasingly distributed generation. Nevertheless, massive deployment of EVs faces great challenges due to: (i) high cost; (ii) limited range due to limited capacity of energy conversion and storage devices, e.g., battery or fuel cell; and (iii) limited number of charging points. Similarly, current energy infrastructure and markets are not ready for a prompt transition to renewable generation and decentralized operations.   

WPT has recently emerged as an promising approach to overcome the aforementioned drawback of EVs deployment \cite{SLi15,CMi16,Hata16,Patil18,Hata18,Machura19}. Especially, the concept of dynamic wireless charging, enabled through wireless charging lanes, help extend traveling ranges of EVs, while giving convenience to EV owners, since no stop and cables are required for charging EVs. Therefore, dynamic wireless charging creates a mutual relation between transportation and energy systems, in which EV serves as a bridge. To further facilitates that mutual relation, this research investigates the bidirectional WPT between EVs and the so-called WCDLs, i.e., EVs are able to not only get charging from WCDLs, but also discharge to WCDLs. 

The EV--WCDL bidirectional energy exchange is attractive due to its capability of on-the-fly charging and discharging -- a great option that brings the ultra-mobility, convenience and comfort for users, where no stop and plug-in cables are required. Furthermore, EV fleets can serve as a super-flexible and clean resource for providing a wider range of ancillary services to the grid. In areas with deep penetration of renewables, e.g., California, USA, or Kyushu, Japan, curtailments on renewable power generation were made to guarantee the supply-demand balance, even with some types of energy storage systems \cite{Yabe19}. In such situations, charging or discharging from a large number of EVs could help reduce the curtailed amount and diminish the so-called duck curve \cite{Yabe19}, where ancillary services can be provided by EV fleets not only at noon when they are parked at homes or offices but also in the morning and in the afternoon when they are moving on roads. As a result, both transportation and energy systems can become low-carbon emission systems.  

From the energy system perspective, a WCDL supplemented with renewable sources (e.g., solar, wind, etc.) along road sides can be regarded as a prosumer who can both produce and consume energy. Likewise, an EV with on-site storage systems (e.g., battery, supercapacitor, etc.) can also be regarded as a prosumer. Thus, energy exchange between an EV group and a WCDL is in fact energy trading between prosumers, of which innovative market models have recently been studied, for instance the so-called peer-to-peer (P2P) energy market. A number of works in the recent literature has been devoted to investigate P2P energy systems, see e.g., \cite{Sorin19,Baroche19,Sousa19,MorstynP2P19b,Tushar19,Alam19,MorstynP2P18,Werth18,NLiu17}. However, in all of those works, the problems of how to select parameters of prosumers' cost functions and how to tune them if the derived energy transactions are unsuccessful or unsatisfied have not been investigated. 

To the author's best knowledge, this research is the first to study the P2P bidirectional energy trading between EVs and WCDLs. Moreover, this research contributes the following to the literature. 
\begin{itemize}
	\item P2P bidirectional energy trading between EVs and WCDLs as an incentive mechanism for EVs to provide ancillary service for the grid, e.g., DR responses.
	\item Analytical solution for the optimal P2P energy market clearing problem, by which the negotiation between EVs and a WCDL is conducted through consensus algorithms enhanced with a privacy-preserving mechanism to avoid exposing private parameters of EVs and the WCDL. 
	\item A fully decentralized method to select parameters in the cost functions of EVs and the WCDL to achieve successful P2P energy transactions with expected energy price and energy amounts. 
\end{itemize}

The rest of this paper is organized as follows. In Section \ref{prob}, system description is given, and problems are formulated. Consequently, a decentralized P2P energy trading mechanism for EVs and the tuning of cost function parameters for the WCDL and EVs are proposed in Section \ref{energy}. The illustrating simulations are presented in Section \ref{app}. Lastly, the paper is summarized in Section \ref{sum}.


\section{System Description} 
\label{prob}

\subsection{Wireless Charging-Discharging Lane}
\label{wcdl}

In this research, it  is assumed that the resonant IWPT technology is used for the WCDL, where coils are placed under the WCDL, and the other coils are attached under the chassis or in the wheels of EVs (see, e.g. \cite{Hata18}, for WPT between wireless charging lanes and EVs with in-wheel motors and coils). 

	\begin{figure}[htpb!]
		\centering
		\includegraphics[scale=0.6]{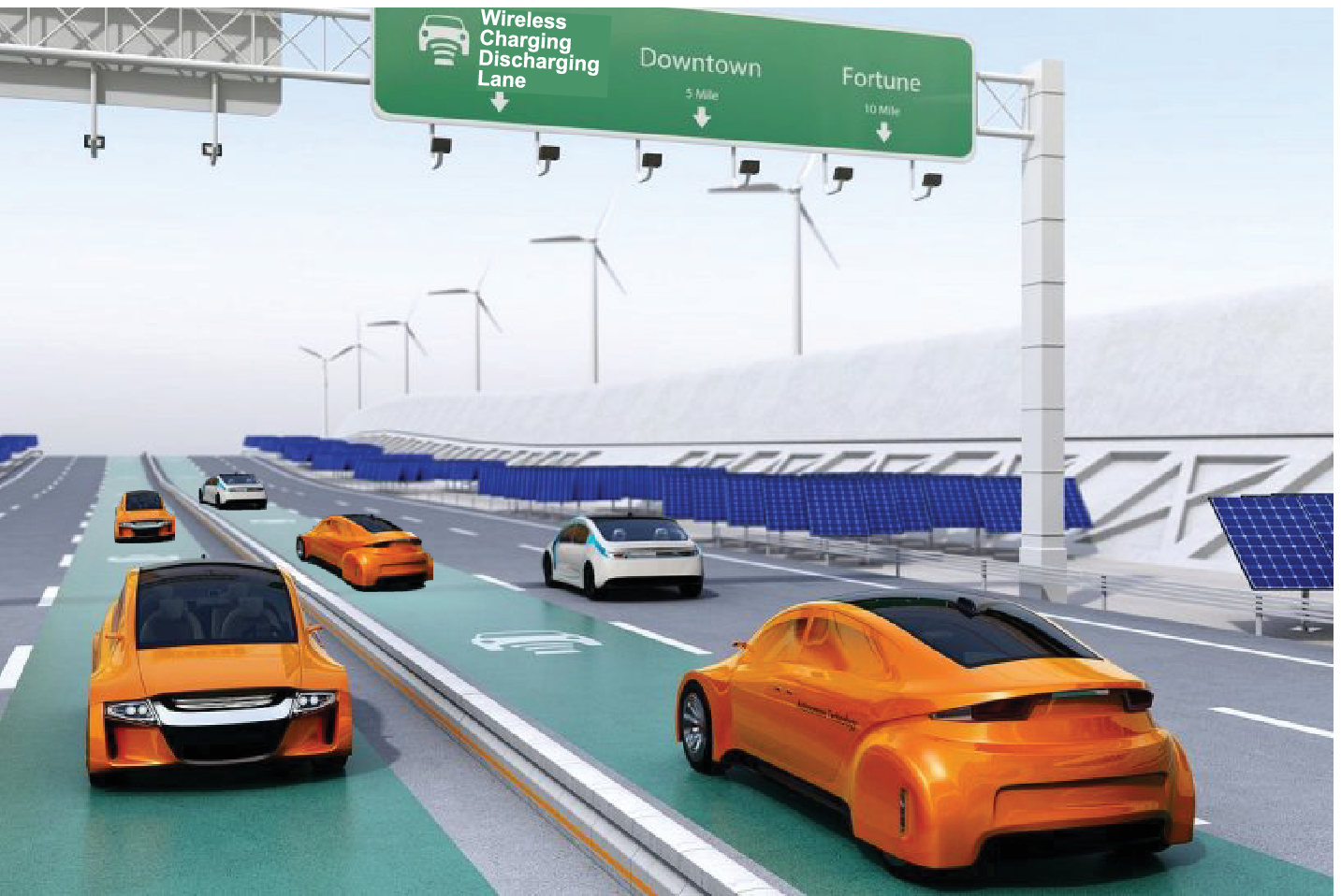}
		\includegraphics[scale=0.6]{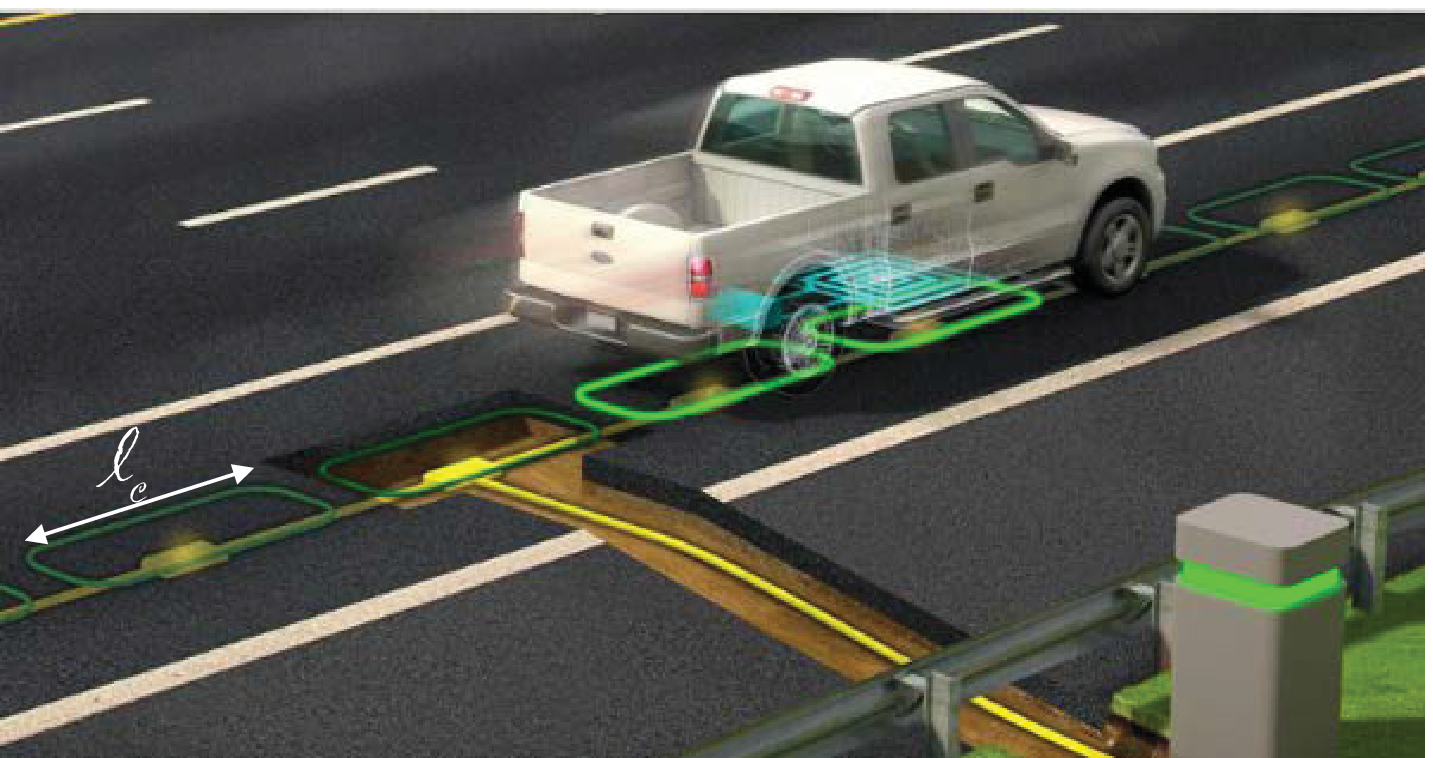}
		\caption{Illustration for EVs moving on wireless charging-discharging lanes (upper), and under-road wireless charging-discharging segments (lower).}
		\label{wireless_coil}
	\end{figure}
  
Denote the number of underground wireless power transceivers by $n_{\rm c}$ and the length of each charging-discharging segment (assumed the same length for all segments) in the WCDL by $\ell_{\rm c}$. When the $i$th EV is charged from the WCDL, the energy it receives is calculated by 
\begin{equation}
	\label{energy-wcl-1}
	E_{c,i} = P_{r} \eta_{d,r} \eta_{c,i} n_{i} \frac{\ell_{c}}{v_{wpt}},
\end{equation}
where $E_{c,i}$ is the received energy by EV $i$; $P_{r}$ is the rated power of each transmitting segment; $\eta_{d,r}$ is the wireless discharging efficiency of the segments; $\eta_{c,i}$ is the charging efficiency of EV $i$; $n_i (\leq n_{c})$ is the number of charging segments that the $i$th EV passes through; and $v_{wpt}$ is the designated velocity on the WCDL. 
A similar formula can be derived when the $i$th EV is discharged to the WCDL as follows,
\begin{equation}
	\label{energy-wcl-2}
	E_{d,i} = P_{EV,i} \eta_{d,i} \eta_{c,r} n_{i} \frac{\ell_{c}}{v_{wpt}},
\end{equation}
where $E_{d,i}$, $P_{EV,i}$, $\eta_{d,i}$ are the discharged energy, discharged power, and wireless discharging efficiency, of EV $i$, respectively; $\eta_{c,r}$ is the wireless charging efficiency of the segments. 

\subsection{Issues on Wireless Energy Exchange for EVs}
\label{sub-probs}

In order for enabling P2P energy exchange between EVs and WCDLs, the P2P market clearing mechanism, which is the main issue, needs to be derived. This P2P energy trading can also be regarded as a mechanism to incentivize EV owners for actively participating in DR programs. Hence, a novel approach will be proposed in Section \ref{energy} to address it. The negotiation between EVs and a WCDL is supported by a proper information and communication infrastructure, assumed readily available.  

An additional issue arises on the coordination of multiple EVs, e.g. when they switch between normal lanes and WCDLs. This problem can be suitably dealt with by platoon formation control methods which have been extensively investigated in the literature as a solution to improve the smoothness of traffic flows and energy saving for the whole vehicle group. 
A well-established framework to study formation control problems is MAS (see, e.g. \cite{Olfati-Saber:2007,Ren:2007,Oh15,Soni17,Hu17}), where each vehicle is cast as an agent. It should be noted that for vehicle formation control, the inter-vehicle information such as relative position and relative speed are most essential.  Therefore, the edge (i.e., the vehicle-to-vehicle or agent-to-agent) dynamical evolution is very important. Recently, several works \cite{Zelazo11,Zelazo13,Zelazo18,Zeng16,Nguyen-TAC:2017,Nguyen-TAC:2018,Chowdhury18,Nguyen-VPPC19} have studied MASs using edge dynamics, by which the formation control problems are converted to equivalent stabilization problems at the origin which is much easier to deal with than the consensus problems on manifolds. However, none of the works in \cite{Zelazo11,Zelazo13,Zelazo18,Zeng16,Nguyen-TAC:2017,Nguyen-TAC:2018,Chowdhury18} considered the formation control problems of vehicles under the changes on speeds of vehicles and on road lanes. In our recent studies \cite{Nguyen-ISCS:20,Nguyen-IFAC:20}, both uncertainty and disturbance in formation control of autonomous EVs have been dealt with. To make the current paper focused, details on formation control of EVs will not be presented.

\section{EV-WCDL Decentralized Privacy-Preserving Optimal Trading}
\label{energy}

In this section, a bidirectional trading mechanism between EV and WCDL prosumers is proposed. 
As mentioned in Section \ref{sub-probs}, the P2P energy trading between EVs and WCDL owners can be considered as an incentive mechanism for EV owners to join in DR programs. This is especially useful to flatten steep ramps (up and down) on the net load curve, which could occur around noon and in the evening when renewable outputs are very high and very low, respectively.   
Therefore, WCDL owners do not set fixed energy price, but instead let EV owners negotiate the energy trading price and amount to encourage them charge or discharge in advance of their plans. It is worth emphasizing that the charging/discharging of EVs through WCDLs considered in this paper does not mean to completely replace the conventional charging/discharging of parked EVs (at homes, offices, etc.), but instead an alternative solution to it. 

There would be multiple WCDLs and many EVs that could exchange energy with the others, however, different from stationay prosumers such as households with rooftop solar panels, EVs are mobile prosumers. Therefore, EVs owners would choose the closest WCDLs (to their routes) for energy trading, i.e., each EV sells or buys energy to only one WCDL. Hence, in the following, we formulate and solve the P2P energy trading between $n$ EVs and one WCDL.

\subsection{Characterization of Optimal Solution}

Each EV prosumer (EV owner) has an objective function, assumed to be quadratic and convex, as follows,
\begin{equation}
\label{EV-func}
f_{{\rm V},i}(E_{{\rm V},i})=a_{{\rm V},i}  E_{{\rm V},i}^2 +b_{{\rm V},i} E_{{\rm V},i}
\end{equation}
where $a_{{\rm V},i}>0$ and $b_{{\rm V},i}>0$ are constant coefficients known only by EV prosumer $i$, $E_{{\rm V},i}$ is the traded energy of EV $i$. 
Likewise, a WCDL prosumer (WCDL owner) has a cost function, also assumed to be quadratic and convex, 
\begin{equation}
\label{wcdl-func}
f_{{\rm L}}(E_{{\rm L}})=a_{{\rm L}}E_{{\rm L}}^2 + b_{{\rm L}}E_{{\rm L}} 
\end{equation}
where $a_{{\rm L}}>0$ and $b_{{\rm L}}>0$ are constant coefficients known only by the WCDL prosumer, $E_{{\rm L}}$ is the traded energy of the WCDL. 
Thus, the optimization to be solved for the P2P energy trading between EVs and a WCDL is
\begin{subequations}
\label{p2p-opt}
\begin{align}
\min \; &  \sum_{i=1}^n f_{{\rm V},i}(E_{{\rm V},i}) + f_{{\rm L}}(E_{{\rm L}}) \\
\label{balance-constraint}
\mathrm{s.t.} ~ & E_{{\rm V},i} + E_{{\rm L},i} = 0, \\
\label{uti-constraint}
& \underline{E}_{{\rm L}} \leq E_{{\rm L}} = \sum_{i=1}^n E_{{\rm L},i} \leq \overline{E}_{{\rm L}}, \\
\label{energy-constraint}
& \underline{E}_{{\rm V},i} \leq E_{{\rm V},i} \leq \overline{E}_{{\rm V},i}.
\end{align}
\end{subequations}

When EVs are charged by the WCDL, $E_{{\rm V},i}>0$, $E_{{\rm L},i}<0$, hence $\underline{E}_{{\rm V},i}=0$, $\bar{E}_{{\rm L}}=0$. 
Likewise, as EVs are discharged to the WCDL, $E_{{\rm V},i}<0$, $E_{{\rm L},i}>0$, and $\bar{E}_{{\rm V},i}=0$, $\underline{E}_{{\rm L}}=0$. For conciseness, in the following only the scenario of EV charging is presented, and the case of EV discharging can be obtained in a similar manner. 

The optimization problem for EV-WCDL cooperation when EVs are charged is as follows. 
\begin{subequations}
\label{p2p-opt-1}
\begin{align}
\min \; &  \sum_{i=1}^n f_{{\rm V},i}(E_{{\rm V},i}) + f_{{\rm L}}(E_{{\rm L}}) \\
\label{balance-constraint-1}
\mathrm{s.t.} ~ & E_{{\rm V},i} + E_{{\rm L},i} = 0, \\
\label{uti-constraint-1}
& \underline{E}_{{\rm L}} \leq E_{{\rm L}} = \sum_{i=1}^n E_{{\rm L},i} \leq 0, \\
\label{energy-constraint-1}
& 0 \leq E_{{\rm V},i} \leq \bar{E}_{{\rm V},i}.
\end{align}
\end{subequations}
The Lagrangian associated with \eqref{p2p-opt-1} is
\small
\begin{align*}
		\tL = & \, \sum_{i=1}^{n}f_{{\rm V},i}(E_{{\rm V},i}) + f_{{\rm L}}(E_{{\rm L}}) - \sum_{i=1}^{n}\lambda_i(E_{{\rm V},i} + E_{{\rm L},i})  \\ 
		& - \mu_{{\rm L},1}(E_{{\rm L}}-\underline{E}_{{\rm L}}) + \mu_{{\rm L},2}E_{{\rm L}} \\
		& - \sum_{i=1}^{n}\hat{\mu}_{{\rm V},i}E_{{\rm V},i} + \sum_{i=1}^{n}\check{\mu}_{{\rm V},i}(E_{{\rm V},i}-\bar{E}_{{\rm V},i}), 
\end{align*}
\normalsize
where $\lambda_i$, $\mu_{{\rm L},1} \geq 0$, $\mu_{{\rm L},2} \geq 0$, $\hat{\mu}_{{\rm V},i} \geq 0$, $\check{\mu}_{{\rm V},i} \geq 0$ are Lagrange multipliers associated with the constraints \eqref{balance-constraint-1}--\eqref{energy-constraint-1}. Next, the following assumptions, which are reasonable in practical situations, are employed. 
\begin{itemize}
	\item[{\bf A1:}] Successful trading for the WCDL and all EVs. 
	\item[{\bf A2:}] Lower bound $\underline{E}_{{\rm L}}$ of WCDL selling energy and upper bound $\bar{E}_{{\rm V},i}$	of EV buying energy are sufficiently small and large, respectively. 
\end{itemize}

\begin{rem}
The above assumptions are employed to simplify the characterization of optimal solutions of \eqref{p2p-opt-1}. Later, in Section \ref{learning}, a method will be introduced to guarantee successful trading between EVs and the WCDL, and to satisfy the constraints  \eqref{uti-constraint-1}--\eqref{energy-constraint-1}, i.e., to satisfy both assumption {\bf A1} and assumption {\bf A2}. 
\end{rem}

Because the cost functions of EVs and the WCDL are assumed as in \eqref{EV-func} and \eqref{wcdl-func} and all constraints are linear, the mathematical programming \eqref{p2p-opt-1} is convex. Therefore, KKT conditions are necessary and sufficient for \eqref{p2p-opt-1}, which read as follows.
\begin{subequations}
\label{kkt}
\begin{align}
\label{kkt-1}
\left. \frac{\partial f_{{\rm V},i}(E_{{\rm V},i})}{\partial E_{{\rm V},i}} \right|_{E_{{\rm V},i}^{\ast}} - \lambda_i - \hat{\mu}_{{\rm V},i} + \check{\mu}_{{\rm V},i} &= 0, \\
\label{kkt-2}
\left. \frac{\partial f_{{\rm L}}(E_{{\rm L}})}{\partial E_{{\rm L},i}} \right|_{E_{{\rm L},i}^{\ast}} - \lambda_i - \mu_{{\rm L},1} + \mu_{{\rm L},2} &= 0, \\
\label{kkt-3}
E_{{\rm V},i}^{\ast} + E_{{\rm L},i}^{\ast} &= 0, \\
\mu_{{\rm L},1}\left(\sum_{i=1}^{n}E_{{\rm L},i}^{\ast}-\underline{E}_{{\rm L}}\right) = 0, ~ \mu_{{\rm L},2}E_{{\rm L},i}^{\ast} &= 0, \notag \\
\label{kkt-4}
\hat{\mu}_{{\rm V},i}E_{{\rm V},i}^{\ast} = 0, ~ \check{\mu}_{{\rm V},i}(E_{{\rm V},i}^{\ast}-\bar{E}_{{\rm V},i}) &= 0,
\end{align}
\end{subequations}  
where $E_{{\rm V},i}^{\ast}$ and $E_{{\rm L},i}^{\ast}$ are optimal values of $E_{{\rm V},i}$ and $E_{{\rm L},i}$, respectively. Then, assumption {\bf A1} leads to $\hat{\mu}_{{\rm V},i}=0$ and $\mu_{{\rm L},2}=0$, while assumption {\bf A2} implies that $\check{\mu}_{{\rm V},i}=0$ and $\mu_{{\rm L},1}=0$. Thus, \eqref{kkt} becomes
\begin{subequations}
\label{kkt-new}
\begin{align}
\label{kkt-new-1}
2a_{{\rm V},i}E_{{\rm V},i}^{\ast} + b_{{\rm V},i} - \lambda_i &= 0, \\
\label{kkt-new-2}
2a_{{\rm L}}\sum_{i=1}^{n}E_{{\rm L},i}^{\ast} + b_{{\rm L}} - \lambda_i &= 0, \\
\label{kkt-new-3}
E_{{\rm V},i}^{\ast} + E_{{\rm L},i}^{\ast} &= 0. 
\end{align}
\end{subequations}  
Equation \eqref{kkt-new-2} reveals that all the energy prices $\lambda_i$ for individual P2P trading between the WCDL and one EV are the same. Denote this unique price by $\lambda$. 
Next, dividing both sides of \eqref{kkt-new-1} by $a_{{\rm V},i}$, both sides of \eqref{kkt-new-2} by $a_{{\rm L}}$, summing them up and utilizing \eqref{kkt-new-3}, we obtain 
\begin{align}
\label{ld}
0 &= \sum_{i=1}^{n}\frac{b_{{\rm V},i}}{a_{{\rm V},i}} + \frac{b_{{\rm L}}}{a_{{\rm L}}} - \lambda \left( \sum_{i=1}^{n}\frac{1}{a_{{\rm V},i}} + \frac{1}{a_{{\rm L}}} \right), \notag \\
\Leftrightarrow 
\lambda &=  \left. \left( \sum_{i=1}^{n}\frac{b_{{\rm V},i}}{a_{{\rm V},i}} + \frac{b_{{\rm L}}}{a_{{\rm L}}} \right) \right/ \left( \sum_{i=1}^{n}\frac{1}{a_{{\rm V},i}} + \frac{1}{a_{{\rm L}}} \right).
\end{align}
Accordingly, the optimal energy to be traded by the WCDL and each EV are as follows,
\begin{equation}
\label{opt-energy}
\begin{aligned}
 \sum_{i=1}^n E_{{\rm L},i}^{\ast} = E_{{\rm L}}^{\ast} &= \frac{1}{2a_{{\rm L}}}\left( \lambda  - b_{{\rm L}} \right),  \\
E_{{\rm V},i}^{\ast} &= \frac{1}{2a_{{\rm V},i}} \left( \lambda  - b_{{\rm V},i} \right).
\end{aligned}
\end{equation}

\subsection{Decentralized Privacy-Preserving Negotiation of P2P Market Clearing Price}

It is obvious from \eqref{ld} that the P2P market clearing price between the WCDL and EVs is calculated using information from all of them. Nevertheless, each EV is only communicated with the WCDL for energy trading, hence a mechanism to attain \eqref{ld} in a decentralized manner is needed. This is achievable by using consensus algorithms for MASs, such as the following. 

Let the WCDL and EVs run a consensus algorithm with variables $x_{0}$ (for the WCDL) and $x_{i}$ (for EV $i$), whose initial values are set to be:
\begin{equation}
	\label{mas-init}
	\scalebox{1}{$
	\begin{aligned}
		x_{0}(0) &= \left[\frac{b_{{\rm L}}}{a_{{\rm L}}},\frac{1}{a_{{\rm L}}}\right]^T,  \\
		x_{i}(0) &= \left[\frac{b_{{\rm V},i}}{a_{{\rm V},i}},\frac{1}{a_{{\rm V},i}}\right]^T, i=1,\ldots,n. 
	\end{aligned}
	$}	
\end{equation}
At time step $k \geq 0$, the WCDL and each EV communicate to run the following consensus algorithm,
\begin{equation}
	\label{consensus-law}
	x_{i}(k+1) = a_{ii}x_{i}(k) + \sum_{j \in \mathcal{N}_{i}}{a_{ij}x_{j}(k)},i=0,1,\ldots,n,
\end{equation}
where $0<a_{ij}<1 \, \forall \, j \in \mathcal{N}_{i}$, $0<a_{ii}<1$ are constant parameters satisfying $\displaystyle \sum_{j=0}^{n}a_{ij}=1 \, \forall \, i=0,\ldots,n$. There are multiple ways to choose $a_{ij}$, e.g. the Metropolis weights \cite{Nguyen-TSG17}. 
Then it can be proved, follows the standard proof for average consensus in the literature (see, e.g. \cite{Olfati-Saber:2007}, \cite{Ren:2007}), that all variables $x_i$ reach the average consensus vector $x_{{\rm ave}} = [x_{{\rm ave},1},x_{{\rm ave},2}]^T$, as $k \rightarrow \infty$, where 
\begin{equation}
\label{ave}
x_{{\rm ave},1} \triangleq \frac{\displaystyle \sum_{i=1}^{n}\frac{b_{{\rm V},i}}{a_{{\rm V},i}} + \frac{b_{{\rm L}}}{a_{{\rm L}}} }{n+1}, ~
x_{{\rm ave},2} \triangleq \frac{\displaystyle \sum_{i=1}^{n}\frac{1}{a_{{\rm V},i}} + \frac{1}{a_{{\rm L}}}}{n+1}.
\end{equation}
As such, the P2P market clearing price $\lambda$ is computed by the WCDL and each EV as follows, 
\begin{equation}
	\label{lambda}
	\lambda = \frac{x_{{\rm ave},1}}{x_{{\rm ave},2}}. 
\end{equation} 

As seen from \eqref{consensus-law}, the initial values of the WCDL and EVs are exchanged, therefore their private parameters $a_{{\rm L}},b_{{\rm L}}$ and $a_{{\rm V},i},b_{{\rm V},i}$ are exposed, which is a critical privacy issue that they do not want. To clear this concern, several approaches can be employed to secure the WCDL-EV information exchange, which can be categorized into encrypted and non-encrypted approaches. For the former, Paillier additive homomorphic cryptosystem is currently one of the most used algorithm for public key cryptography (see, e.g. \cite{Ruan19}). For the latter, a few studies have been conducted to obtain secured consensus algorithms that converge exactly to the average of initial values (see, e.g. \cite{YMo17}). While the former can provide better privacy guarantee, its computational complexity is higher, hence longer computational time. Thus, there is always a tradeoff between privacy and computation overhead for secure consensus algorithms. 

It is worth emphasizing that {\it decentralized cryptosystem} is still a hard problem. For example, the work in \cite{Ruan19} required an assumption that each agent has at least a trustable neighboring agent who does not try to infer the other agent's initial condition. On the other hand, the masking approach in \cite{YMo17} necessitated the non-overlapping neighboring sets between agents, therefore in star networks, such as that in the current research, can only guarantee the privacy of the center node (the WCDL in the current research), but cannot protect the privacy of the other nodes (EVs in the current research). To this end, derivation of a decentralized privacy-preserving algorithm for consensus problem, which is applicable to any network topology and uses non-conservative assumptions, needs much more works, hence should be considered in a separated study.    

For the current research, if we assume that the WCDL owner has a limited computability that prevents it from trying to infer private parameters of many EVs communicated to it for P2P energy trading, i.e., initial values of EVs in \eqref{mas-init}, then the masking approach in \cite{YMo17} can be utilized. 
Each peer (whether the WCDL or an EV) creates a masked state 
\begin{equation}
	\label{masked-state}
	\tilde{x}_i(k) = x_i(k) + [w_{i,1}(k),w_{i,2}(k)]^T,
\end{equation} 
in which $w_{i,1}(k)$ and $w_{i,2}(k)$ are random noises generated by:
\begin{equation}
	\label{mask}
	w_{i,\ell}(k) = \left\{ 
	\begin{aligned}
		& \zeta_{i,\ell}(0), \quad & {\rm if} \; k=0, \\
		& \alpha_{i}^{k}\zeta_{i,\ell}(k)-\alpha_{i}^{k-1}\zeta_{i,\ell}(k-1), \quad & {\rm otherwise,}
	\end{aligned}
	\right.
\end{equation}
for $\ell=1,2$, where $\zeta_{i,\ell}(k)$ are Gaussian random variables independently generated by each peer $j$ from a standard normal distribution, i.e., a normal distribution with mean $0$ and variance $1$; and $0<\alpha_{j}<1$ is a constant. Then the secure consensus algorithm is given by,
\begin{equation}
	\label{secure-consensus}
	x_{i}(k+1) = a_{ii}\tilde{x}_{i}(k) + \sum_{j \in \mathcal{N}_{i}}{a_{ij}\tilde{x}_{i}(k)},i=0,1,\ldots,n.
\end{equation}
It was proved in \cite{YMo17} the privacy-preserving consensus algorithm \eqref{secure-consensus} converges exactly to the average vector $x_{{\rm ave}}$ shown in \eqref{ave}. 

Note that $\alpha_{i}$ are distinct for $i=0,1,\ldots,n$, hence the noise $w_i(k)$ generation is completely independent (fully decentralized) for the considering peers.

\section{Selection of Cost Function Parameters for Desired EV-WCDL P2P Energy Trading}
\label{learning}

As seen in Section \ref{energy}, the P2P energy trading between the WCDL and EVs strongly depends on their cost functions, more specifically their cost function parameters $a_{\rm L},b_{\rm L}$, and $a_{{\rm V},i},b_{{\rm V},i}$. Nevertheless, how to set the values of those parameters for deriving expected energy transaction price and amount is {\it ad hoc} for the WCDL and each EV owner, and has not been addressed in the literature. Therefore, in this section a cooperative learning strategy is proposed to tune cost function parameters of the WCDL and EVs for attaining desired energy transactions, based on the analytical solution \eqref{ld}--\eqref{opt-energy} of the P2P optimal clearing problem \eqref{p2p-opt-1}.   

\subsection{Fully Decentralized Setting of Cost Function Parameters} 

Since $E_{{\rm L}}^{\ast}<0$ and $E_{{\rm V},i}^{\ast}>0 \; \forall \; i=1,\ldots,n$, it is obtained from \eqref{opt-energy} that 
\begin{equation}
	\label{ld-bounds}
	b_{{\rm V},i}^{\max} < \lambda < b_{\rm L},
\end{equation}
where $b_{{\rm V},i}^{\max} \triangleq \displaystyle \max_{i=1,\ldots,n} b_{{\rm V},i}$. 
Therefore, the WCDL and EVs need to set their parameters $b_{\rm L}$ and $b_{{\rm V},i}$ properly to obtain a desired energy trading price. Here, it is proposed that each EV and the WCDL select its range of expected trading price, denoted by $[\underline{\lambda}_i,\overline{\lambda}_i],i=0,1,\ldots,n$, where the subscript $0$ represents the WCDL. Consequently, these price ranges will be exchanged between EVs and the WCDL to obtain a common range of price for all. This cooperative negotiation procedure follows standard consensus algorithms similarly to that in \eqref{consensus-law}, where lower and upper bounds of EV and the WCDL prices are updated by such consensus algorithms, as follows. 
\begin{equation}
\begin{aligned}
	\label{consensus-law-price}
	\underline{\lambda}_i(k+1) = a_{ii}\underline{\lambda}_i(k) + \sum_{j \in \mathcal{N}_{i}}{a_{ij}\underline{\lambda}_i(k)}, \\
	\overline{\lambda}_i(k+1) = a_{ii}\overline{\lambda}_i(k) + \sum_{j \in \mathcal{N}_{i}}{a_{ij}\overline{\lambda}_i(k)},
\end{aligned}	
\end{equation}
for $i=0,1,\ldots,n$, where $a_{ij}, i,j=0,\ldots,n$ are the same as for \eqref{consensus-law}.  
Note that no secure algorithm is needed here because EVs and the WCDL need to know exactly the price range of the other. 

After reaching consensus on the price range, denoted by $[\underline{\lambda},\overline{\lambda}]$, EVs and WCDL need to set their parameters to assure successful P2P energy transactions, i.e., to satsify \eqref{ld-bounds}. To do so, 
EVs and WCDL choose their parameters by the following rule, 
\begin{equation}
	\label{b-selection}
	b_{{\rm V},i} \in \left[\underline{\lambda},\frac{1}{2}(\underline{\lambda}+\overline{\lambda})\right), ~
	b_{\rm L} \in \left(\frac{1}{2}(\underline{\lambda}+\overline{\lambda}),\overline{\lambda}\right]
\end{equation}
which ensure $\displaystyle \max_{i=1,\ldots,n} b_{{\rm V},i} < b_{\rm L}$ and $\lambda \in [\underline{\lambda},\overline{\lambda}]$. 
Then it can be easily shown that $\lambda < b_{\rm L}$ by utilizing \eqref{ld}. 
Next, to guarantee that $b_{{\rm V},i}^{\max} < \lambda$, we substitute $\lambda - b_{{\rm V},i}^{\max}$ into \eqref{ld} to obtain the following condition,
\begin{align*}
0 <  \sum_{i=1}^{n}\frac{b_{{\rm V},i} - b_{{\rm V},i}^{\max}}{a_{{\rm V},i}} + \frac{b_{{\rm L}} - b_{{\rm V},i}^{\max}}{a_{{\rm L}}},
\end{align*}
which is true if
\begin{align}
\label{b-cond-1}
& 0 <  \sum_{i=1}^{n}\frac{b_{{\rm V},i}^{\min} - b_{{\rm V},i}^{\max}}{a_{{\rm V},i}} + \frac{b_{{\rm L}} - b_{{\rm V},i}^{\max}}{a_{{\rm L}}}  \notag \\
\Leftrightarrow \quad & (b_{{\rm V},i}^{\max} - b_{{\rm V},i}^{\min})\sum_{i=1}^{n}\frac{1}{a_{{\rm V},i}} <  \frac{b_{{\rm L}} - b_{{\rm V},i}^{\max}}{a_{{\rm L}}},
\end{align}
where $b_{{\rm V},i}^{\min} \triangleq \displaystyle \min_{i=1,\ldots,n} b_{{\rm V},i}$. Due to \eqref{b-selection}, we further obtain the following condition as a sufficiency for \eqref{b-cond-1}, hence for \eqref{ld-bounds}, 
\begin{align}
\label{b-cond}
\frac{\overline{\lambda}-\underline{\lambda}}{2}\sum_{i=1}^{n}\frac{1}{a_{{\rm V},i}} <  \frac{1}{a_{{\rm L}}} \left(b_{{\rm L}} - \frac{\overline{\lambda}+\underline{\lambda}}{2}\right).
\end{align}
Next, let $[\underline{E}_{{\rm L}}, 0)$ and $(0, \overline{E}_{{\rm V},i}]$ be the ranges of desired energy amounts to be traded for the WCDL and EV $i$, as in \eqref{uti-constraint-1} and \eqref{energy-constraint-1}. 
Utilizing \eqref{opt-energy} and \eqref{ld-bounds}, we obtain
\begin{align}
 E_{{\rm V},i}^{\ast} < \frac{1}{2a_{{\rm V},i}} \left( b_{{\rm L}}  - b_{{\rm V},i} \right) \leq \frac{1}{2a_{{\rm V},i}} \left( b_{{\rm L}}  - b_{{\rm V},i}^{\min} \right).
\end{align}
Note that $b_{{\rm L}} \leq \overline{\lambda}$ and $b_{{\rm V},i}^{\min} \geq \underline{\lambda}$, therefore a sufficient condition for attaining $E_{{\rm V},i}^{\ast} < \overline{E}_{{\rm V},i}$ is that
\begin{align}
\label{a-cond-ev}
 \frac{1}{2a_{{\rm V},i}} \left( \overline{\lambda}  - \underline{\lambda} \right) \leq \overline{E}_{{\rm V},i} \; \Leftrightarrow \; \frac{1}{2a_{{\rm V},i}} \leq \frac{\overline{E}_{{\rm V},i}}{\overline{\lambda}  - \underline{\lambda}}.
\end{align}
This gives EVs a way to choose their parameters $a_{{\rm V},i}$ in a completely decentralized manner. 

Now, substituting \eqref{a-cond-ev} into \eqref{b-cond} results in the following condition for $a_{{\rm L}}$ such that \eqref{b-cond} is satisfied,
\begin{align}
\label{a-cond-l-1}
\sum_{i=1}^{n}\overline{E}_{{\rm V},i} <  \frac{1}{a_{{\rm L}}} \left(b_{{\rm L}} - \frac{\overline{\lambda}+\underline{\lambda}}{2}\right) \; \Leftrightarrow \; a_{{\rm L}} < \frac{b_{{\rm L}} - \frac{\overline{\lambda}+\underline{\lambda}}{2}}{\displaystyle \sum_{i=1}^{n}\overline{E}_{{\rm V},i}}.
\end{align}
On the other hand, the following should be satisfied for the WCDL,
\begin{align}
\label{b-cond-3}
 \underline{E}_{{\rm L}} < E_{{\rm L}}^{\ast} = \frac{\lambda - b_{{\rm L}}}{2a_{{\rm L}}} = \frac{1}{2a_{{\rm L}}} \frac{\sum_{i=1}^{n}\frac{b_{{\rm V},i}-b_{{\rm L}}}{a_{{\rm V},i}}}{\sum_{i=1}^{n}\frac{1}{a_{{\rm V},i}}+\frac{1}{a_{{\rm L}}}}.
\end{align}
We have $b_{{\rm V},i}-b_{{\rm L}} \geq b_{{\rm V},i}^{\min}-b_{{\rm L}} \geq \underline{\lambda}  - \overline{\lambda}$, hence the following condition is sufficient for \eqref{b-cond-3},
\begin{align}
\label{b-cond-4}
(\underline{\lambda}  - \overline{\lambda}) \frac{\frac{1}{2a_{{\rm L}}}\sum_{i=1}^{n}\frac{1}{a_{{\rm V},i}}}{\frac{1}{\sum_{i=1}^{n}a_{{\rm V},i}}+\frac{1}{a_{{\rm L}}}} &> \underline{E}_{{\rm L}}  \notag \\
\Leftrightarrow \; 2a_{{\rm L}} + \frac{2}{\sum_{i=1}^{n}\frac{1}{a_{{\rm V},i}}} &> \frac{\underline{\lambda}  - \overline{\lambda}}{\underline{E}_{{\rm L}}}.
\end{align}
Using \eqref{a-cond-ev}, the following is sufficient for \eqref{b-cond-4},
\begin{align}
\label{a-cond-l-2}
2a_{{\rm L}} + \frac{\overline{\lambda}  - \underline{\lambda}}{\sum_{i=1}^{n}\overline{E}_{{\rm V},i}} &> \frac{\underline{\lambda}  - \overline{\lambda}}{\underline{E}_{{\rm L}}} \notag \\
 \Leftrightarrow \; a_{{\rm L}} &> \frac{\overline{\lambda}  - \underline{\lambda}}{2} \left( \frac{-1}{\underline{E}_{{\rm L}}} - \frac{1}{\sum_{i=1}^{n}\overline{E}_{{\rm V},i}} \right).
\end{align}
Note that the range for $a_{{\rm L}}$ specified in \eqref{a-cond-l-1} and \eqref{a-cond-l-2} requires the upper bounds $\overline{E}_{{\rm V},i}$ of traded energy from EVs, which are sent to the WCDL as a part of negotiation procedure.

\begin{algorithm}
	\label{wcdl-ev-p2p}
	\caption{Decentralized P2P Energy Trading for Charging EVs from the WCDL}
	\SetKwInOut{Input}{Input}
	\SetKwInOut{Output}{Output}
	
	WCDL and EVs choose their initial price ranges; 
			
	EVs select maximum energy amounts $\overline{E}_{{\rm V},i}$ to be charged through the WCDL, and send them to the WCDL;
	
	WCDL chooses its upper bound $\underline{E}_{{\rm L}}$ of energy amount to trade with EVs; 

	\% {\it Negotiation of energy trading price range}
	
	\For{$1 \leq k \leq max\_iter$}{
	
	WCDL and EVs run the consensus algorithm \eqref{consensus-law-price};
	
	\If{$k=max\_iter$, or $|\underline{\lambda}_i(k+1)-\underline{\lambda}_i(k)| \leq \epsilon$, $|\overline{\lambda}_i(k+1)-\overline{\lambda}_i(k)| \leq \epsilon \; \forall \; i=0,\ldots,n,$}
		{
			break\;
		}		
	}	
		
	WCDL and EVs obtain the common P2P energy trading price range $[\underline{\lambda},\overline{\lambda}]$;	

	\% {\it Selection of cost function parameters}
	
	WCDL and EVs select $b_{{\rm L}}$ and $b_{{\rm V},i}$ to satisfy \eqref{b-selection}; 
	
	EVs choose $a_{{\rm V},i}$ to satisfy \eqref{a-cond-ev};
	
	WCDL selects $a_{{\rm L}}$ to satisfy \eqref{a-cond-l-1} and \eqref{a-cond-l-2};

	\% {\it Privacy-preserving negotiation of P2P energy trading price}	
	
	\For{$1 \leq k \leq max\_iter$}{
	
	WCDL and EVs run the masked consensus algorithm \eqref{secure-consensus};
	
	\If{$k=max\_iter$, or $\|\tilde{x}_i(k+1)-\tilde{x}_i(k)\|_2 \leq \epsilon \; \forall \; i=0,\ldots,n,$}
		{
			break\;
		}		
	}	
		
	WCDL and EVs compute the P2P energy trading price $\lambda$ by \eqref{lambda};
			
	WCDL and EVs compute the P2P energy trading amount by \eqref{opt-energy};		
	
\end{algorithm}

\subsection{Summary of The Proposed P2P Energy Trading Mechanism}

Denote $max\_iter$ the maximum number of iterations for the consensus algorithms \eqref{consensus-law}, \eqref{secure-consensus}, and \eqref{consensus-law-price}. Let $\epsilon$ be a given small positive number. 
The proposed decentralized P2P energy trading mechanism between the WCDL and EVs is summarized in Algorithm  \ref{wcdl-ev-p2p}.

\section{Numerical Simulation}
\label{app}

This section aims at demonstrating the proposed P2P energy trading algorithm between the WCDL and EVs. Assume that the rated power by the WCDL is 400kW, the resonant IWPT efficiency is $\eta_{d,r}= 90\%$, the conversion efficiency of the electronic circuit on EVs is $\eta_{c,i}= 95\%$, the total length of wireless charging segments is 3 km, and the speed of EVs on the WCL is $v_{\rm wpt}=50$ km/h (which is the limit on most urban roads in Japan), then the maximum energy that one EV can get from the WCDL, computed in \eqref{energy-wcl-1}, is 20.52kWh. Here, the number of EVs is first assumed to be 50. 

\begin{figure}[htpb!]
		\centering
		\includegraphics[scale=0.4]{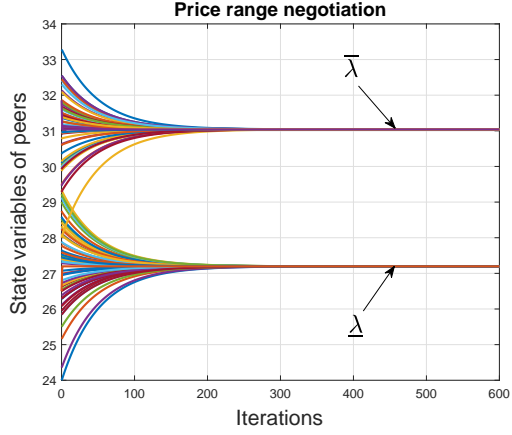}
		\caption{Negotiation of P2P energy price between the WCDL and EVs.}
		\label{ev_wcdl_price_range}
	\end{figure}

As shown in Algorithm \ref{wcdl-ev-p2p}, the WCDL and EVs first set their initial price range for the negotiation. It is noted that the feed-in tariff (FIT) in Japan for the fiscal year 2020 is 21 JPY/kWh for under 10kW solar generation \cite{Meti20}.  Therefore, it is assumed here that the WCDL initially set its price range to be $[24,28]$ JPY/kWh to incentivize EVs, whereas EVs expect a higher price with their initial lower and upper bounds of price ranges randomly selected around 27 JPY/kWh and 31 JPY/kWh. Then, utilizing the consensus algorithm \eqref{consensus-law-price}, the negotiation between WCDL and EVs is depicted in Figure \ref{ev_wcdl_price_range}. It is obtained that $\underline{\lambda}=27.2$ JPY/kWh, and $\overline{\lambda}=31.04$ JPY/kWh. 

\begin{figure}[htpb!]
		\centering
		\includegraphics[scale=0.4]{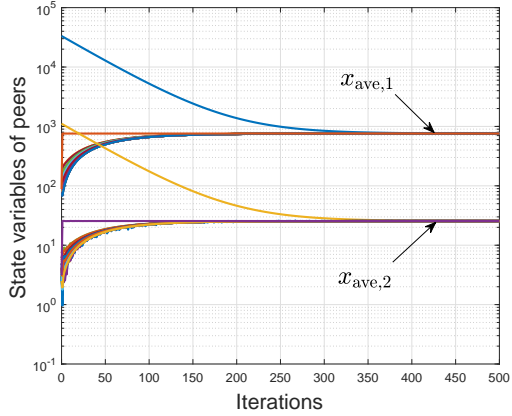}
		\caption{Privacy-preserving consensus of peers.}
		\label{ev_wcdl_masked_css}
	\end{figure}

\begin{figure}[htpb!]
		\centering
		\includegraphics[scale=0.4]{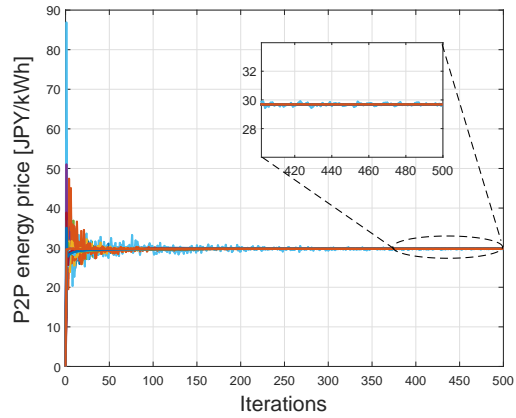}
		\caption{Privacy-preserving P2P trading energy price along the negotiation.}
		\label{ev_wcdl_masked_price}
	\end{figure}

\begin{figure}[htpb!]
		\centering
		\includegraphics[scale=0.4]{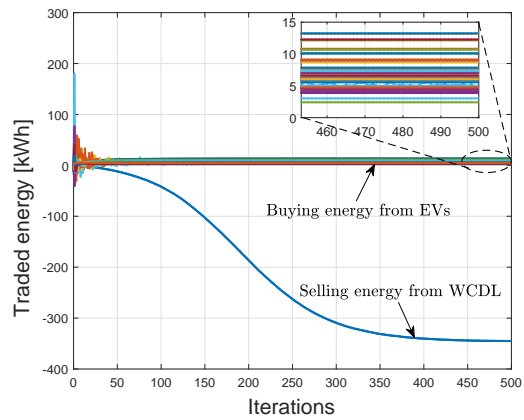}
		\caption{P2P traded energy of WCDL and EVs.}
		\label{ev_wcdl_energy}
	\end{figure}

Next, EVs set their maximum amounts of traded energy $\overline{E}_{{\rm V},i} = 15$kWh, and send to the WCDL. Then the WCDL set $\underline{E}_{{\rm L}} = -700$kWh.  
Consequently, following \eqref{b-selection}, WCDL choose its $b_{{\rm L}}$ to be 30, while EVs randomly select their $b_{{\rm V},i}$ between 27.2 and 29.12. In the next step,  EVs randomly choose their parameters $a_{{\rm V},i}$ such that they satisfy \eqref{a-cond-ev}, which in this case reads $a_{{\rm V},i} \geq 0.128$. For the WCDL, condition \eqref{a-cond-l-2} is always satisfied here, because the right hand side is negative. On the other hand, condition \eqref{a-cond-l-1} says $a_{{\rm L}}<0.0012$, hence it is chosen to be $0.0009$. Afterward, WCDL and EVs run the masked consensus algorithm \eqref{secure-consensus} to derive the P2P market clearing price $\lambda$, whose results are shown in Figure \ref{ev_wcdl_masked_css}--\ref{ev_wcdl_energy}. It can be observed that even in the presence of added noises, state variables of WCDL and EVs still converge to their averages, and hence, their ratio, i.e., the P2P market clearing price converge exactly to the optimal solution \eqref{ld}, as depicted in Figure \ref{ev_wcdl_masked_price}. Moreover, all EVs are successfully traded with the WCDL, as exhibited in Figure \ref{ev_wcdl_energy}.

\begin{figure}[htpb!]
		\centering
		\includegraphics[scale=0.4]{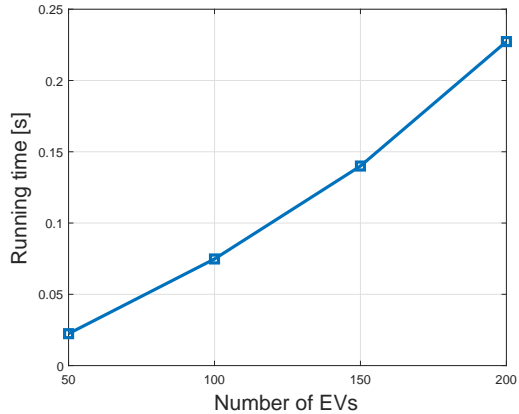}
		\caption{Running time of the proposed EV--WCDL P2P energy trading algorithm (without communication time between EVs and WCDL).}
		\label{ev_wcdl_scale}
	\end{figure}

Finally, the scalability of the proposed P2P energy trading algorithm is tested, where the number of EVs is increased from 50 to 100, 150, and 200. It is noted that all the results presented in Section \ref{energy} and Section \ref{learning} are analytical, hence the running time of the proposed decentralized P2P energy trading algorithm for EVs and WCDL depends on that of the consensus protocols and communication time between EVs and WCDL. Here, the latter is ignored, and only the former is checked, whose results are plotted in Figure \ref{ev_wcdl_scale}. It can be observed that the running time is increased almost linearly with system size, thus the proposed algorithm is scalable well.  

\section{CONCLUSION}
\label{sum}

A decentralized P2P energy trading algorithm has been proposed in this paper for energy exchange between EVs and a WCDL. Analytical formulas for the P2P market clearing price and optimal energy trading amounts have been obtained, based on which a method has been introduced to properly select the cost function parameters of both the EVs and the WCDL such that all peers successfully trade with desired energy price and energy amounts. It is remarkable that this method is also analytical, hence no iterative procedure is needed to tune such parameters. Further, a privacy-preserving approach has been employed to protect peers from private information leak. The proposed algorithm performance and scalability are well verified through a test case.

\section*{Acknowledgement}
The author's research is financially supported by JSPS Kakenhi Grant Number 19K15013.


\bibliographystyle{plain} 
\bibliography{References}

\end{document}